\documentstyle[psfig]{EuroPhys}
\input EuroMacr
\input epsf.sty
\begin{document}
\euro{}{}{}{1998}
\Date{26 May 1998}
\shorttitle{A. Lambrecht et al. Generating photons pulses with an 
oscillating cavity}
\title{Generating photon pulses with an oscillating cavity}
\author{A. Lambrecht\inst{1}, M.T. Jaekel\inst{2} \And S. Reynaud\inst{1}}
\institute{
\inst{1}Laboratoire Kastler Brossel \thanks{%
Laboratoire de l'Ecole Normale Sup\'erieure et de l'UPMC associ\'e au CNRS}, 
Universit\'{e} Pierre et Marie Curie, case 74\\
4 place Jussieu, 75252 Paris, France\\
\inst{2}Laboratoire de Physique Th\'eorique de l'Ecole Normale Sup\'erieure 
\thanks{Laboratoire du CNRS associ\'e \`a l'Ecole Normale Sup\'erieure et 
\`a l'Universit\'e Paris Sud}\\ 24 rue Lhomond, 75231 Paris, France}
\rec{April 1998}{May 1998}
\maketitle

\begin{abstract}
We study the generation of photon
pulses from thermal field fluctuations through opto-mechanical coupling to
a cavity with an oscillatory motion. Pulses are regularly spaced and become 
sharp
for a high finesse cavity. 
\end{abstract}

Photon production from vacuum field fluctuations reflected onto moving
boundaries is now a well known phenomenon \cite{Moore70,Fulling76}. For a single
moving mirror motion-induced radiation is quite small since its magnitude is 
determined by
the mirrors velocity over the speed of light while the velocity 
of any macroscopic mirror is restricted to the order of
the sound velocity.
A way to enhance the effect is to use two mirrors
which form a Fabry-P\'{e}rot cavity. 
Motion-induced effects then take profit of resonance enhancement \cite{Jaekel91} 
and
the number of photons radiated by the cavity is shown to approach orders of 
magnitude of
experimental demonstrations \cite{Lambrecht96}. 
The resonance enhancement takes place when motional photons
are emitted into optical cavity modes. This occurs when the mechanical 
oscillation frequency is adjusted such
as to be a multiple of optical resonance frequencies. 
Motional radiation may be
interpreted as resulting from the dephasings undergone by the field as it is
reflected on the moving mirrors. 
Inside a cavity, the field dephasings add up continously over
successive reflections and may therefore become
large. The total dephasing is determined by an effective velocity obtained as 
the
product of the physical velocity with the finesse which gives roughly
the number of roundtrips of a photon inside the cavity. When the effective
velocity approaches the speed of light photon production concentrates in sharp
pulses \cite{Law94,Cole96,Lambrecht98}.

The aim of the present letter is to study the pulse shaping effect for itself. 
Indeed the opto-mechanical coupling
between an oscillatory motion of a scatterer and field fluctuations produces a 
temporal
redistribution of field fluctuations which leads to the emission of pulses. 
Here we will consider a cavity moving in thermal fluctuations. We expect this 
situation 
to allow for the observation of the 
pulse shaping effect in a much simpler experimental configuration than for 
vacuum fluctuations. 
Incidentally, the same computation will give the temperature which would be 
necessary 
in order to see pulse shaping for quantum vacuum.

\begin{figure}[tbp]
\centerline{\psfig{figure=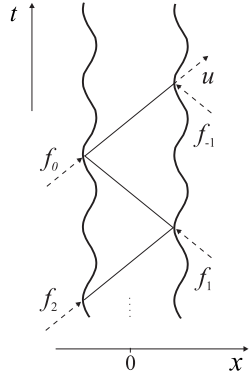,height=5cm}}
\caption{Space-time diagram showing the construction of the dephasings 
$f_p$ of light rays reflecting back and forth in an oscillating cavity. 
The oscillating lines represent the trajectories of mirrors moving in phase. 
The straight lines making a 45$^{\circ}$
light rays since the speed of light is set to unity. }
\label{fig1}
\end{figure}

To evaluate the motional radiation we have to evaluate the field correlation
function at the output of the cavity. The main lines of the derivation are
explained in detail in \cite{Lambrecht98} and are
not repeated here. Neglecting polarisation effects, the electromagnetic field is 
modelled as 
the sum of two scalar components counterpropagating in a two-dimensional 
space-time and
thus defined as functions of the light cone variables $u=t-x$ and $v=t+x$
(the speed of light is set to unity).
The two-dimensional model gives valid predictions in
four-dimensional space-time provided that the waist of the cavity modes
is much smaller than the mirrors surface. The dephasings
felt by the fields as they undergo multiple reflections onto the mirrors are
described by functions $f_{p}(u)$ representing the optical transformation of
various input rays into a given output ray as shown in figure \ref{fig1}. 

For a cavity with two mirrors at rest at positions $\pm \frac{L}{2}$ those
functions are simply given by $f_{p}(u)=u-pL$. The mechanical cavity length $%
L$ is measured as a time of flight of photons from one mirror to the other.
The particular ray $f_{-1}$ represents the field directly scattered at the
outer side of the cavity by the first encountered mirror. The problem of
finding the functions $f_{p}$ when mirrors are moving is difficult in
the general case. When large dephasings are obtained, the functions
corresponding to single reflections cannot be simply added but have
rather to be composed. We will use here the solution obtained in \cite
{Lambrecht98} for a harmonic motion which produces a very 
efficient opto-mechanical coupling between fields and the moving cavity.

We consider a harmonic oscillation at a frequency $\Omega =K\frac{\pi }{L}$ of 
order $K$ 
with respect to the lowest optical cavity mode.  
The amplitude of the mirrors motion is governed by a rapidity parameter $%
\alpha $ the hyperbolic tangent of which gives the maximum mechanical
velocity $v = {\rm th}\alpha$ normalised to the speed of light.  
The two mirrors are described either by the same rapidity for $K$ odd (global 
translation
mode of the cavity) or by
opposite values for $K$ even (breathing mode of the cavity). The mirrors 
mechanical 
velocity is small
compared to the speed of light so that the parameter $\alpha $ is much
smaller than unity. The calculation of the functions $f_{p}$ is then reduced to 
a
composition of homographic functions and solved in an analytical manner as 
\begin{eqnarray}
e^{i\Omega f_{p}(u)} &=&\frac{a_{p}e^{i\Omega u}+b_{p}}{b_{p}^{*}e^{i\Omega
u}+a_{p}^{*}}  \nonumber \\
a_{p}=(-i)^{Kp}{\rm ch}p\alpha  &\qquad &b_{p}=i^{2K+1}(-i)^{Kp}{\rm sh}%
p\alpha   \label{homogr}
\end{eqnarray}
This means that the rapidity $\alpha $ adds up to $p\alpha $ under the effect of 
$p$ successive reflections. Mathematically this is due to the group structure of 
homographic functions \cite{Lambrecht98}. 
The quantity $p\alpha $ has its order of magnitude
determined by an effective rapidity 
\begin{equation}
\alpha _{{\rm eff}}=2\frac{\alpha }{\rho }  \label{eff}
\end{equation}
where $\frac{1}{\rho }$ is the mean number of roundtrips experienced by the
field before it leaves the cavity.
For a high finesse cavity the effective velocity can reach large values
approaching the speed of light. In fact a threshold for parametric
oscillation exists at $\alpha _{{\rm eff}}=1$. At this value the energy
density diverges and above threshold the system should show self-sustained
oscillations. This behaviour which has been discussed in the limiting case
of vanishing temperature \cite{Lambrecht98} is independent of the field
temperature and restricts the predictions of the present calculations to the
range $\alpha _{{\rm eff}}<1$.

The energy density emitted by the cavity is then computed as the field
correlation function at two coinciding points with the help of a point
splitting regularization \cite{Fulling76}. The expression for the
energy density emitted by the cavity through mirror 2 is
found to be 
\begin{eqnarray}
e_{u} &=&\frac{\hbar R_{2}}{48\pi }\left\{ \Omega ^{2}\left( f_{-1}^{\prime
2}-1\right) +\theta ^{2}f_{-1}^{\prime 2}\right\}   \nonumber \\
&+&\frac{\hbar T_{1}T_{2}}{48\pi }\sum_{n\geq 0}r^{2n}\left\{ \Omega
^{2}\left( f_{2n}^{\prime 2}-1\right) +\theta ^{2}f_{2n}^{\prime 2}\right\} 
\nonumber \\
&+&\frac{\hbar T_{2}^{2}R_{1}}{48\pi }\sum_{n\geq 0}r^{2n}\left\{ \Omega
^{2}\left( f_{2n+1}^{\prime 2}-1\right) +\theta ^{2}f_{2n+1}^{\prime
2}\right\}   \nonumber \\
&+&\frac{\hbar T_{2}}{8\pi }\theta ^{2}\sum_{n\geq 0}r^{n+1}\frac{%
f_{-1}^{\prime }f_{2n+1}^{\prime }}{{\rm sh}^{2}\frac{\theta }{2}%
(f_{-1}-f_{2n+1})}  \nonumber \\
&-&\frac{\hbar T_{1}T_{2}}{16\pi }\theta ^{2}\sum_{n\neq m\geq 0}r^{n+m}%
\frac{f_{2n}^{\prime }f_{2m}^{\prime }}{{\rm sh}^{2}\frac{\theta }{2}%
(f_{2n}-f_{2m})}  \nonumber \\
&-&\frac{\hbar T_{2}^{2}R_{1}}{16\pi }\theta ^{2}\sum_{n\neq m\geq 0}r^{n+m}%
\frac{f_{2n+1}^{\prime }f_{2m+1}^{\prime }}{{\rm sh}^{2}\frac{\theta }{2}%
(f_{2n+1}-f_{2m+1})}  \nonumber \\
\theta  &=&\frac{2\pi k_{{\rm B}}T}{\hbar }\qquad r=e^{-2\rho }=\sqrt{R_1 R_2}
\label{ecav}
\end{eqnarray}
The coefficient $r$ determines the attenuation factor of the field on a
single cavity round-trip and can also be related to the loss parameter $\rho 
$. $R_{i}$ and $T_{i}$ are the reflection and transmission coefficients of
the two mirrors $i=1,2$. They are related through unitarity conditions $%
R_{i}+T_{i}=1$. $\theta $ is the field temperature $T$ in frequency units
where $k_{{\rm B}}$ is the Boltzmann constant. 

The output field correlations already obtained for a cavity oscillating in
quantum vacuum are recovered at the limit of vanishing temperature. In
particular, the hyperbolic sine functions together with the prefactor of $%
\theta ^{2}$ simplify to a function typical of vacuum fluctuations. In the
high temperature regime $\theta \gg \Omega $ in contrast, the terms
depending on hyperbolic sine functions are vanishingly small and may be
ignored. A simple interpretation may therefore be proposed in this
classical limit where the energy density is obtained as the result of a
stretching or tightening of the energy flow lines associated with the time
variation of the dephasing. The presence of additional terms at
low temperature means that this classical interpretation has a restricted
range of validity.

\begin{figure}
\hbox to\textwidth{\vbox{\hbox to 
0.5\textwidth{\psfig{figure=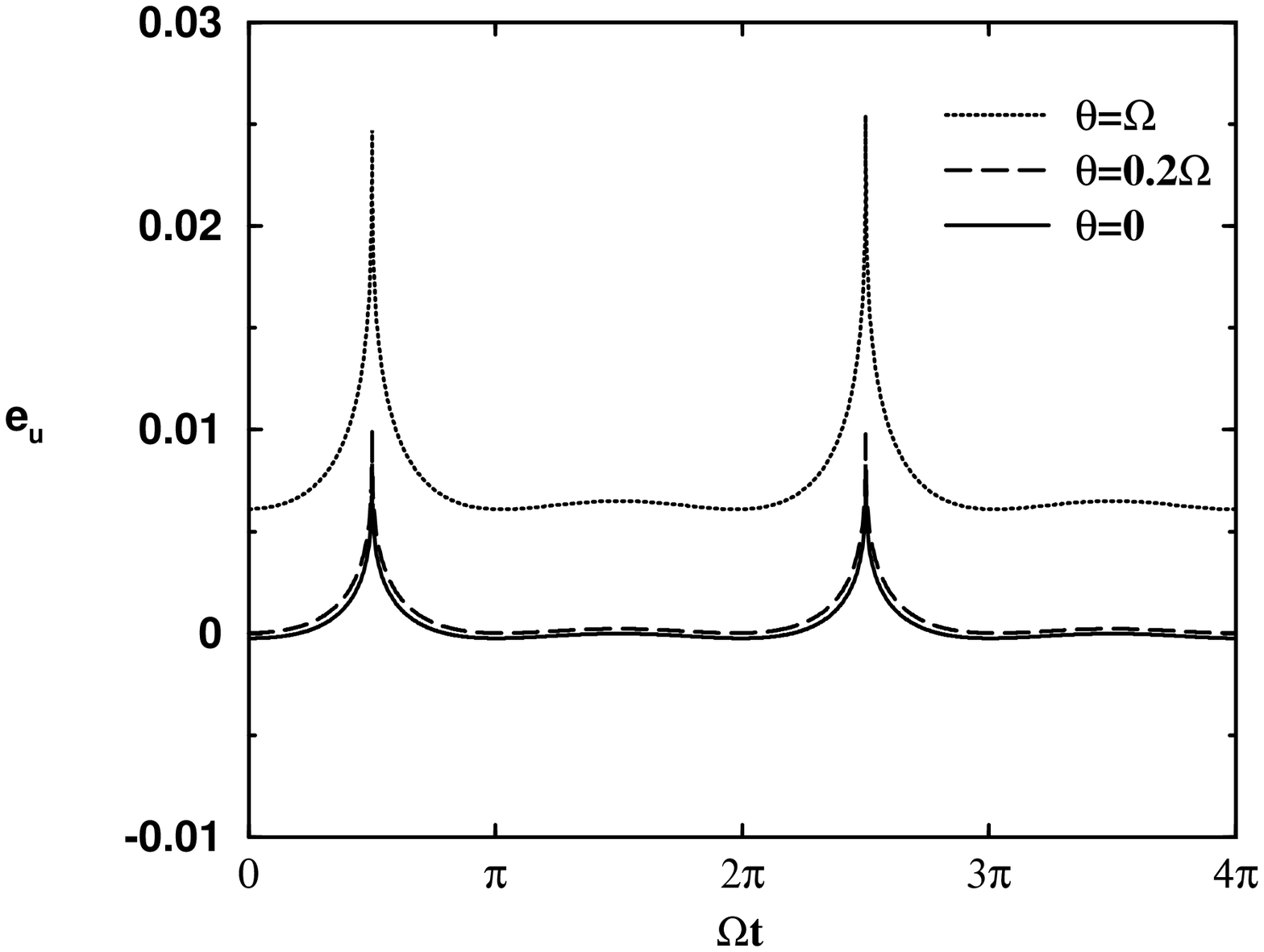,height=6cm}\hfill}}
\vbox{\hbox to 0.5\textwidth{\psfig{figure=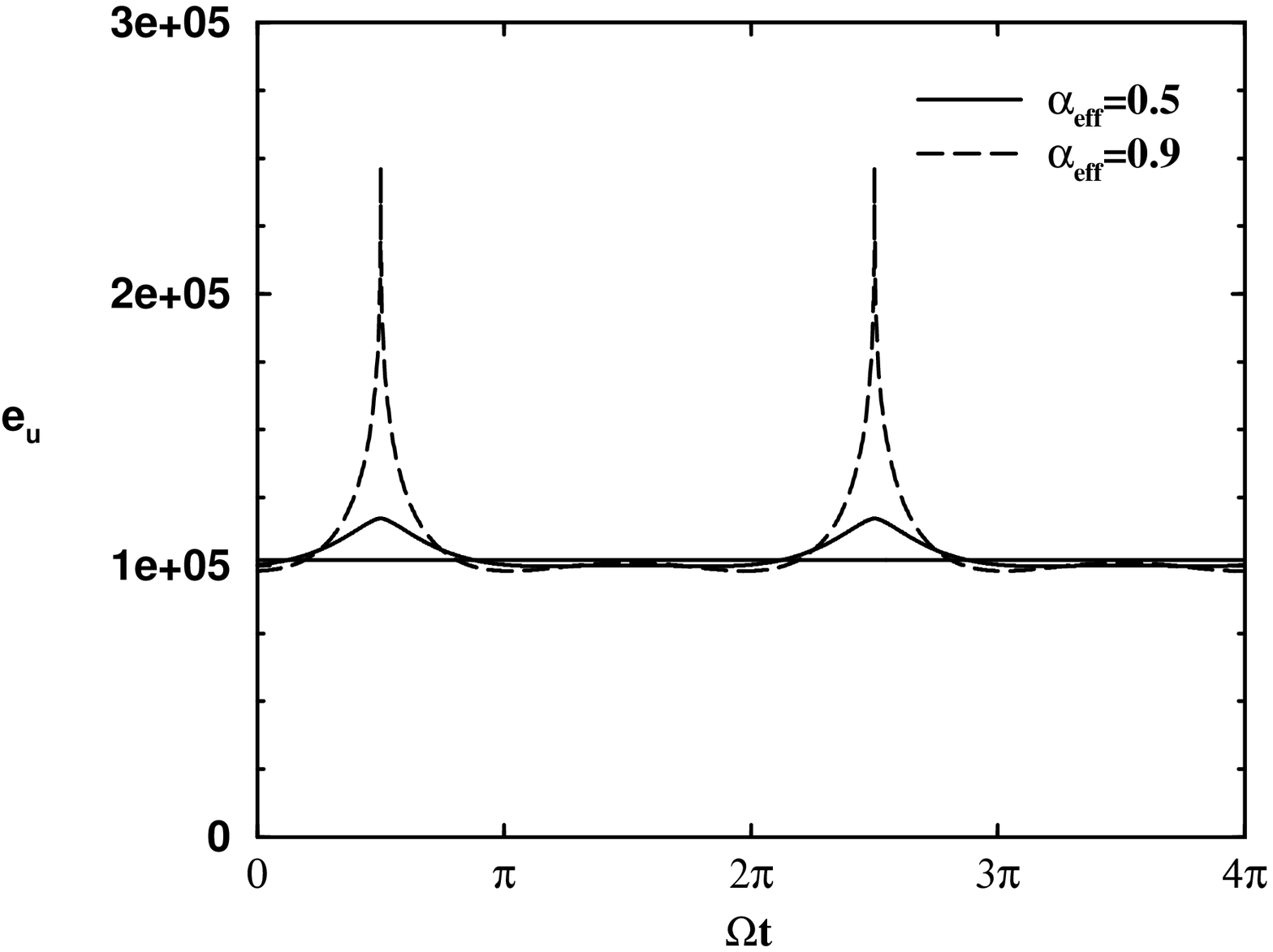,height=6cm}\hfill}}}
\caption{Energy density emitted outside the cavity as a function of time at very
low temperatures. Effective rapidity and attenuation factor are 
$\alpha _{{\rm eff}}=r=0.9$.
The solid line corresponds to quantum vacuum, while the two other plots 
correspond to temperatures $\theta=0.2\Omega$ and $\theta=\Omega$.}
\label{fig2}
\caption{Energy density emitted outside the cavity as a function of time in the 
high
temperature domain ($\theta = 3924\Omega$ corresponding to $T=300$K for 
$\Omega=10$GHz) 
for two different effective rapidities 
$\alpha _{{\rm eff}}=0.5$ (dashed line) and $\alpha_{\rm eff} =0.9$ (solid 
line). 
The attenuation factor is $r=0.9$.
The straight line corresponds to the thermal background directly reflected
by the cavity.}
\label{fig3}
\end{figure}

The resulting energy density emitted by the cavity as a function of time is
plotted in figures \ref{fig2} and \ref{fig3} respectively in the low and high 
temperature domain. For simplicity we
have supposed mirror 1 to be perfectly reflecting ($T_{1}=0$). Both
figures show that pulses emerge from the cavity at regularly spaced times.
The pulses are superimposed to the thermal background which is reflected
back by the cavity and represents the level with respect to which they 
have to be distinguished. The corresponding energy density $\frac{\hbar
\theta ^{2}}{48\pi }$ is the same as for a motionless cavity.
Figure \ref{fig2} shows the result for quantum vacuum
compared to fields of temperature $\theta =0.2\Omega $ and $\theta=\Omega$. 
It allows to evaluate the temperature which should be
maintained in order to measure the pulse shaping effect due to quantum
vacuum. This temperature should be of the order of $\frac{1}{10}$ of the
mechanical oscillation frequency. Already at $\theta \sim \Omega $ pulse
shaping would mainly be produced by the transformation of thermal photons.
For a mechanical oscillation frequency of $\Omega =10$GHz, the temperature
should be as low as $T=10$mK for demonstrating dissipative motional effect of
vacuum.

On the other hand, figure \ref{fig3} clearly shows that pulse shaping in a
thermal field is much like pulse shaping in vacuum for an oscillatory motion
of the cavity. The pulses shown for two different rapidities $\alpha _{{\rm %
eff}}=0.5$ and $\alpha _{{\rm eff}}=0.9$ become
sharper and higher when the effective rapidity is increased towards the
threshold value $\alpha _{{\rm eff}}=1$. The efficiency of pulse shaping
depends only on the value of the effective rapidity and not on temperature.
However the orders of magnitude of the effect are much more favorable in the
high temperature domain. The pulse maxima are growing in the same manner as
the thermal background for increasing temperature. At the same time the
number of photons per pulse increases continously. The regularly
spaced radiation pulses could be more easily detected since they contain
much more photons than at low temperature. If the effective rapidity
is close to its threshold value the pulses emerge
sufficiently from the thermal background to be distinguished. 

We may get a precise idea of the order of magnitude of the motional
radiation effect by evaluating the total energy in a single pulse and 
estimating the number of photons per pulse. To this aim, we perform
the straightforward analytical integration over one oscillation period of
energy density (\ref{ecav}).  
To avoid lengthy expressions, we give the total
energy only in the limiting case of a high cavity
finesse where the orders of magnitude of pulse shaping are more favorable.
We write the total emitted energy $E$ and the energy inside the cavity ${\cal 
E}$ for an 
integration time
equal to a period $\frac{2\pi }{\Omega }$ of the mechanical oscillation 
\begin{eqnarray}
E &=&\frac{\hbar \theta^{2}}{12 \Omega}+\frac{\hbar \Omega }{6}\frac{%
\rho \alpha ^{2}}{\rho ^{2}-\alpha ^{2}}F(\theta)+\frac{\hbar \Omega }{%
6}\alpha ^{2}(1+\theta^{2})  \nonumber \\
{\cal E} &=&\frac{\hbar K\theta^{2}}{24 \Omega}+\frac{\hbar \Omega K}{%
24}\frac{\alpha ^{2}}{\rho ^{2}-\alpha ^{2}}F(\theta)  \nonumber \\
F(\theta) &=&1+\frac{\theta^{2}}{\Omega^2}\left( 1-24\sum_{l=1}^{\infty 
}\frac{1}{%
{\rm sh}^{2}(2\pi Kl\theta/\Omega)}\right)
\label{Etot}
\end{eqnarray}
The intracavity energy ${\cal E}
$ is measured with respect to the Casimir energy which is the reference
level obtained for a motionless cavity at zero temperature. For both
expressions, the terms independent of $\alpha $ correspond to the background
of thermal fluctuations integrated over one oscillation period. The other
terms are proportional to $\alpha ^{2}$ and therefore associated with the
opto-mechanical coupling between the moving cavity and the field
fluctuations. The third term in $E$ which is proportional to $\alpha ^{2}$
but independent of $\rho $ describes the effect of reflection on the
first encountered mirror. Evaluations in the present letter are restricted
to situations below the parametric oscillation threshold $\alpha =\rho /2$.
The most favorable orders of magnitude are thus obtained when $\alpha $
approaches this threshold value 
\begin{eqnarray}
E_{{\rm thr}} &=&\frac{\hbar\theta^{2}}{12\Omega}+\frac{\hbar \Omega 
}{18}\rho F(\theta)+\frac{\hbar \Omega }{24}\rho ^{2}(1+\theta^{2})  \nonumber 
\\
{\cal E}_{{\rm thr}} &=&\frac{\hbar K\theta^{2}}{24\Omega}+\frac{\hbar \Omega 
K}{72}F(\theta)  
\label{Emax}
\end{eqnarray}
For the sake of simplicity, we may focus our attention on the high
temperature limit where $F(\theta)$ has a simple quadratic dependence
on temperature 
\begin{equation}
\frac{\theta} {\Omega}\gg 1\quad \Rightarrow \quad F(\theta)\approx 
\frac{\theta^{2}}{\Omega^2}
\end{equation}
Finally, the order of magnitude of the number of photons is
evaluated by dividing energies by $\hbar \Omega $ and then multiplying the
result by $2$ since each quantum of mechanical excitation $\Omega $ gives
rise to the creation of about $2$ photons \cite{Lambrecht98}. With these
simplifying assumptions, we may now discuss orders of magnitude of the
motional radiation effects.

In principle, the detection of the pulses should be possible by measuring
the time-dependent energy density either outside or inside the cavity.
Inside the cavity, motion-induced pulses may contain up to $\frac{1}{3}$ of
the number of stationnary thermal photons. For an oscillation frequency of $%
\Omega =10$GHz and for the global translation mode $K=3$ the cavity contains
more than $10^{6}$ motion-induced photons at room temperature. These photons are 
easily
distinguishable from the background since they are gathered in sharp pulses,
propagating back and forth the cavity. The energy density then reaches values as 
large as 
the background level multiplied by the cavity finesse. 

The situation is less favorable for the field radiated outside the cavity.
The reason is that the cavity finesse has to be chosen high enough to
compensate the smallness of the oscillation velocity, at the best of the
order of the sound velocity, compared to the velocity of light. As a
consequence, the probability for a photon to escape from the cavity is small. 
This is why the ratio of motion-induced photons to background
photons goes down to $\frac{2\rho }{3}$ outside the cavity. We may still
take profit of the fact that they are emitted in sharp pulses to
distinguish them. In fact, the contrast of pulse energy density is of the
order of $1$ with respect to background energy density. Alternatively, we may
discriminate motion-induced photons in the spectral domain since they are
emitted at cavity resonance frequencies \cite{Lambrecht98}, for example at
frequencies $\Omega /3,2\Omega /3,4\Omega /3,\ldots $ for the mode $K=3$,
whereas background fluctuations have a standard thermal spectrum outside the
cavity since they are obtained from input fluctuations through a mere
reflection on the outer side of the cavity.

There remains the question whether there may be an appreciable number of
motion-induced photons per pulse outside the cavity. In the high temperature
limit we deduce from (\ref{Emax}) the number of photons $N$ per 
pulse as a function of cavity finesse and temperature 
\begin{equation}
\frac{\theta}{\Omega}\gg 1
\quad \Rightarrow \quad N\sim \frac{\rho }{9}\frac{\theta^{2}}{\Omega^2}
\end{equation}
The fact that energy increases as squared temperature may thus
counterbalance the smallness of the cavity transmission. At room temperature
each pulse may contain about $20$ motion-induced photons for $%
\rho =\alpha /2\sim 10^{-5}$ that is for material velocities of the order of
sound velocity. Continously measuring the temporal variation
of the energy density should reveal the presence of these pulses superimposed 
to the background.

Pulse shaping from a 
thermal field appears quite similar to the effect in quantum vacuum. Its
efficiency is essentially the same in both cases and depends
only on the effective rapidity. 
However, an interesting difference arises for the mode $K=1$. Photon production 
into this mode
is suppressed by the cavity bandwidth when the cavity is moving in vacuum 
(cf. equation (\ref{Etot}) for $\theta=0$).
This is due to the fact that it implies photons to be emitted
at frequencies close to zero which are filtered by the cavity. In a thermal 
field 
in contrast the number of photons diverges at low frequencies which 
counterbalances
the effect of the cavity bandwidth. As a result photon production occurs for 
$K=1$ 
with the same
order of magnitude than for any other mode.
 
Pulse shaping from field fluctuations is much more easily detectable in a 
thermal field 
than in quantum vacuum. First a single pulse contains many
more photons when it is generated from a thermal field rather than from
vacuum. Second the experimental setup is considerably simpler 
because there is no need to reach and maintain the
extremely low temperatures which would be necessary to demonstrate pulse
shaping from vacuum fluctuations.

\stars

\end{document}